\documentstyle[prd,aps,epsf]{revtex}

\newcommand{\sikib}{\begin{eqnarray}}
\newcommand{\sikie}{\end{eqnarray}}
\newcommand{\pdt}{\partial_t}
\newcommand{\pdx}{\partial_x}

\newcommand{\px}{\phi}
\newcommand{\vpx}{\varphi}
\newcommand{\ps}{\tilde{\phi}}
\newcommand{\al}{\alpha}
\newcommand{\be}{\beta}
\newcommand{\ep}{\varepsilon}
\newcommand{\om}{\omega}
\newcommand{\omf}{\omega'}

\newcommand{\sqal}{\sqrt{\al}}
\newcommand{\la}{\lambda}
\newcommand{\ka}{\kappa}

\begin{document}

\draft

\twocolumn[\hsize\textwidth\columnwidth\hsize\csname
@twocolumnfalse\endcsname

\title{Black hole radiation with high frequency dispersion}

\author{ Hiromi {\sc Saida}$^a$\footnotemark[3]
         and
         Masa-aki {\sc Sakagami}$^b$\footnotemark[4] }
\address{
  $^a$
  Graduate School of Human and Environmental Studies, Kyoto University,
  Kyoto 606-8501, Japan, \\
  $^b$
  Department of Fundamental Sciences, FIHS, Kyoto University, 
  Kyoto 606-8501, Japan }


\maketitle

\begin{abstract}

We consider one model of a black hole radiation, in which the 
equation of motion of a matter field is modified to cut off high 
frequency modes. The spectrum in the model has already been 
analytically derived in low frequency range, which has resulted in 
the Planckian distributin of the Hawking temperature. On the other 
hand, it has been numerically shown that its spectrum deviates 
from the thermal one in high frequency range. In this paper, we 
analytically derive the form of the deviation in the high frequency 
range. Our result can qualitatively explain the nature of the 
numerically calculated spectrum. The origin of the deviation is 
clarified by a simple discussion. 

PACS number(s): 04.70.Dy, 04.62.+v, 04.70.-s

\end{abstract}


\vskip2pc
]

\footnotetext[3]{E-mail: saida@phys.h.kyoto-u.ac.jp}
\footnotetext[4]{E-mail: sakagami@phys.h.kyoto-u.ac.jp}


\section{Introduction}\label{sec-intro}

There are two important theories in modern physics: quantum 
mechanics and the theory of general relativity. The former 
has been applied to field theory in flat spacetime and marked 
triumphs in understanding the wide range of microscopic phenomena. 
On the other hand, the latter has given us insights into phenomena 
having close relation to gravity. The theory of general relativity, 
which is a classical theory, has drastically changed our 
understandings of space and time. Yet, we have no idea to treat 
the gravity in Planck scale. It is unclear whether the unification 
of quantum mechanics and the theory of general relativity will be 
accomplished in the context of quantum physics. Today, however, we 
do not have any idea superior to the quantum mechanics in treating 
Planck scale physics. It is, therefore, meaningful to study how 
quantum and/or semi-classical effects will be expected in the 
presence of the strong gravity.

The most remarkable discovery including a semi-classical 
gravitational effect is the Hawking radiation\cite{ref-H}, which 
is concluded by treating matter fields on spacetime as quantum 
while a black hole metric as classical. According to this theory, 
a black hole radiates particle flux of a thermal spectrum, whose 
temperature is $\kappa/2\pi$ where $\kappa$ is the surface 
gravity.

The derivetion of the original Hawking radiation relies on the 
extremely high, over Planck scale, frequency modes. These modes 
arise from the extremely high gravitational red shift which the 
radiation undergoes during propagating from the event horizon to the 
asymptotically flat region. It is the investigation of the Hawking 
radiation that sheds light on Planck scale physics.

In the study of this physics, one model of a black hole radiation
\cite{ref-CJ1} \cite{ref-CJ2} \cite{ref-CJ3} has been proposed, which 
makes an assumption on the effects of unknown Planck scale physics as 
follows: the equation of motion (EOM) of matter field is modified to 
have a cut off of the high frequency modes with respect to free fall 
observer.
\footnote[1]{ This model was motivated by the dumb hole\cite{ref-U}, 
which is a hydrodynamical analogue of a black hole radiation. } 
There are two important properties of this model. First one is that 
the modified EOM of matter field violates Lorentz invariance. 
Secondly, the origin of the black hole radiation we consider in this 
paper, which is called the mode conversion,
\footnote[2]{ The phenomenon called mode conversion has already been 
known in plasma physics.\cite{ref-MC1}\cite{ref-MC2}\cite{ref-MC3} } 
is different from that of the original Hawking radiation. We call 
the treatment of the black hole radiation in this paper as 
{\it the mode conversion} ({\it MC}) {\it model} hereafter.

The spectrum of the flux in two dimensional MC model has already 
been numerically calculated\cite{ref-CJ1}. It has been shown that, 
although the spectrum almost agrees with the Planckian 
distributionits, it deviates from the thermal one in the high 
frequency region $\ka \ll \om$, where $\om$ is the energy 
(frequency) of a massless scalar field. On the other hand, an 
analytical calculation\cite{ref-CJ2} has resulted in the thermal 
spectrum, where the analysis has been carried out perturbatively up 
to lowest orders of $\ka$ and $\om$ in the range $\om <\ka$. This is 
appropriate because the derivation of the spectrum in \cite{ref-CJ2} 
does not take the effects of the high frequency range into account.

The purpose of this paper is to extend the analytical calculation in 
\cite{ref-CJ2} to the high frequency range, $\ka<\om$. We obtain the 
form of the distribution function, $N(\om)$, as
  \sikib
    N(\om) \simeq
    \frac{1-\al\om^2/2}
         { \exp[(2\pi\om/\ka)(1-\al\om^2/2)] -1 }
  \nonumber \, ,
  \sikie
which shows the deviation from the thermal spectrum. Here, $\al$ is 
the square of the cut off scale characterizing the new physics in 
Planck scale. This result denotes the same tendency as what 
numerical calculations show. Furthermore, we will give a simple 
explanation of the occurrence of the deviation.

In section \ref{sec-MC}, we explain the MC model\cite{ref-CJ1}. 
Section \ref{sec-spectrum} is devoted to the calculation of the 
spectrum. Lastly, we summarize the results in section \ref{sec-SD}. 

Throughout this paper, we use Planck units, $\hbar = c = G = 1$, 
and restrict our discussion in two dimension for simplicity.


\section{Mode Conversion (MC) model}\label{sec-MC}

We consider the static spacetime with the metric of the form
\cite{ref-CJ1} \cite{ref-CJ2} \cite{ref-CJ3}
  \sikib
    ds^2 = - dt^2 + (\, dx - v(x) \, dt \,)^2 \, .
    \label{eq-1}
  \sikie
It is obvious that the timelike Killing vector field of this 
spacetime is $\pdt$. This spacetime has an event horizon at $x_h$ 
satisfying $v(x_h) = -1$. The surface gravity of this spacetime, 
$\ka$, is calculated to be $\ka=(dv/dx)(x_h)$ which is $1/4M$ for 
the case of the Schwarzschild black hole of mass $M$. Note that the 
world line of $dx - v(x) \, dt = 0$ is normal to the surface, $t=$ 
const, and that it is a geodesic of a free fall observer. The 
velocity of the free fall observer with respect to the rest observer 
is $v(x)$, where we require $v(x) < 0$ and monotone increasing 
$\pdx v(x) > 0$, that is, $v(x \! \to \! \infty) \! \to \! v_0$ 
where $-1<v_0 \leq 0$ by definition. The proper time of this free 
fall observer is $t$ of the coordinate (\ref{eq-1}).

The action of a massless scalar field\cite{ref-CJ1} \cite{ref-CJ2} 
\cite{ref-CJ3} is
  \sikib
    S = \frac{1}{2} \int d^{2}x \sqrt{-\mbox{\rm g}}
        \left[
        \left( {\bf u}(\Phi) \right)^{\ast} {\bf u}(\Phi) -
        \left( {\bf s}(\Phi) \right)^{\ast} {\bf s}(\Phi)
        \right]
        \, ,
    \label{eq-2}
  \sikie
where $\mbox{\rm g} \!=\! \det g_{\mu\nu}$, ${\bf u}$ is a 
unit tangent to the free fall world line and ${\bf s}$ is a unit 
space-like vector which is outward-pointing and orthogonal to 
${\bf u}$. We choose the derivatives of $\Phi$ along ${\bf u}$ 
and ${\bf s}$ as ${\bf u}(\Phi) = u^{\mu} \partial_{\mu} \Phi$ and 
${\bf s}(\Phi) = \hat{F}(s^{\mu} \partial_{\mu}) \Phi$, where 
$\hat{F}(s^{\mu} \partial_{\mu})$ is a function of the differential 
operators, $\partial_{\mu}$, and we require 
$\hat{F}(-s^{\mu} \partial_{\mu}) = -\hat{F}(s^{\mu} \partial_{\mu})$. 
This modification, $\hat{F}(s^{\mu} \partial_{\mu})$, expresses the 
MC model's assumption about the effects of unknown Planck scale 
physics. This leads the EOM of $\Phi$ in the coordinate of 
(\ref{eq-1}) to be 
  \sikib
    \left( \pdt + v(x) \, \pdx \right)
    \left( \pdt + \pdx \, v(x) \right) \Phi =
    \hat{F}^2(\pdx) \Phi \, .
     \label{eq-3}
  \sikie
Hereafter we specify the form of $\hat{F}(\pdx)$ as 
$\hat{F}^2(\pdx) = \pdx^2 + \al \, \pdx^4$, where $\al = 1/k_c^2$ 
and $k_c$ is the cut off scale which characterizes the new physics 
in Planck scale. We set the order of $\al$ is of unity, 
$O(\al) \sim 1$.

\begin{figure}[t]
  \epsfysize=55mm
  \epsfbox{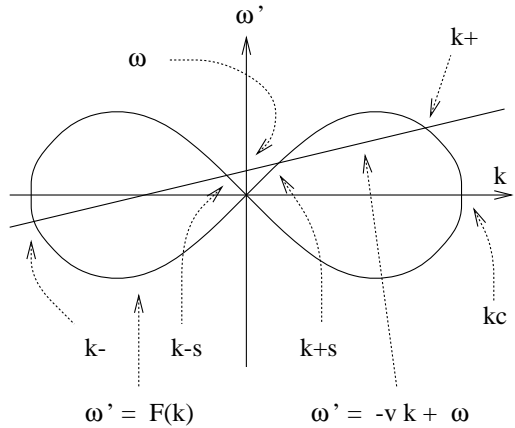}
  \caption{Dispersion relation of our MC model.}
  \label{fig-dispersion1}
\end{figure}

We require that, at least far from the event horizon, $v(x)$ is 
nearly constant enough to require the validity of WKB approximation. 
This means that $\pdx v(x)/k(x) \ll 1$ and 
$\pdx k(x)/k^2(x) \ll 1 \, (k \simeq$ const.$)$ far from the event 
horizon. With setting 
$\Phi \simeq \exp \left( -i \om t + ikx \right)$, the EOM 
(\ref{eq-3}) gives the dispersion relation
  \sikib
    \left\{ 
     \begin{array}{ccl}
       \omf & = & -v k + \om \\
       \omf & = & \pm \sqrt{k^2 - \al k^4} \equiv F(k)
     \end{array}
    \right. \, ,
    \label{eq-4}
  \sikie
where $\omf$ and $\om$ are the frequencies with respect to the free 
fall observer and to the rest observer respectively, since 
${\bf u}(\Phi) = -i (\om - v k)\Phi = -i \omf \Phi$ and 
$\pdt \Phi = -i\om \Phi$. The Killing frequency, $\om$, is conserved 
during time evolution but the free fall frequency, $\omf$, is not. 
Equations (\ref{eq-4}) are the dispersion relation of this MC 
model, which have four mode solutions for fixed $\om$. We call the 
wave numbers of these modes as $k_{-}$, $k_{-s}$, $k_{+s}$ and 
$k_{+}$ in increasing order, as shown in the figure 
\ref{fig-dispersion1}. It is not the signature of $\om$ but of $\omf$ 
according to which we can judge whether the solution is the positive 
frequency mode or the negative frequency one\cite{ref-CJ1}.

\begin{figure}[t]
  \epsfysize=60mm
  \epsfbox{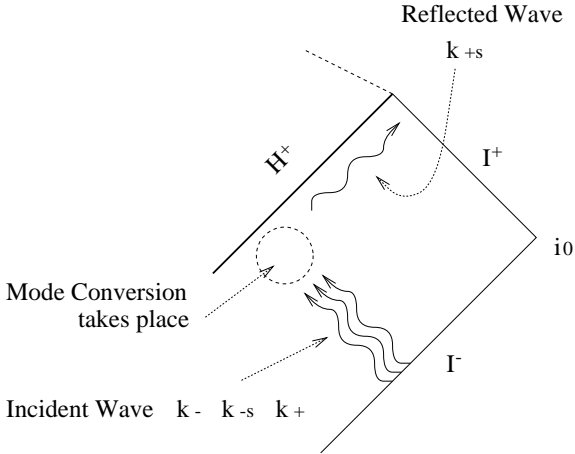}
  \caption{Schematical description of the mode conversion on an 
           asymptotically flat spacetime. I$^{\pm}$ are the future 
           and past null infinity. i$_0$ is the spatial infinity. 
           H$^+$ is the event horizon.}
  \label{fig-mcrad}
\end{figure}

\begin{figure}[t]
  \epsfysize=55mm
  \epsfbox{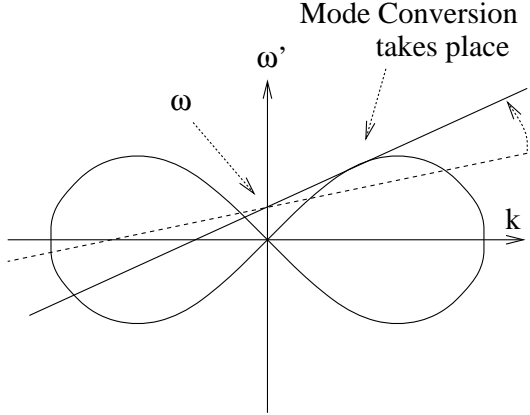}
  \caption{Graphical description of the mode conversion in the graph 
           of dispersion relation. In view of tracing the wave 
           packet of $k_{+s}$ backward in time, it is converted to 
           the wave packet of $k_{-s,\pm}$.}
  \label{fig-dispersion2}
\end{figure}

The origin of the black hole radiation in this model can be 
understood by analyzing a wave packet propagation\cite{ref-CJ1}
\cite{ref-CJ2}\cite{ref-CJ3}. The group velocity of a wave packet 
with respect to the free fall observer and the rest observer are 
expressed as $V'_g = d\omf(k)/dk$ and $V_g = V'_g + v(x)$ 
respectively. In the case of $\om > 0$, we can find easily that 
$k_{-s,\pm}$ modes are of ingoing, while $k_{+s}$ one is of 
outgoing. In order to understand the physics of mode conversion, 
the hydrodynamical analogue of black hole radiation is helpful
\cite{ref-U}\cite{ref-CJ1}. In this model, a fluid flows toward a 
center, on which we consider a wave packet propagation. Here, $v(x)$ 
in (\ref{eq-1}) represents the velocity of the infalling fluid flow. 
As shown in the figure \ref{fig-mcrad}, $k_{+s}$ mode travels 
outward away from the event horizon, and against the infalling flow 
of back ground fluid. In tracing the mode backward in time, it 
approaches toward the event horizon. In the figure 
\ref{fig-dispersion2}, we show how $k_{+s}$ mode moves in the 
diagram of the dispersion relation. Near the event horizon, its 
group velocity, $V'_g(k_{+s})$, eventually coincides with the fluid 
velocity, $v(x)$. Here the mode conversion takes place. In the view 
of our tracing backward in time, the infalling wave packet of 
$k_{+s}$ never reaches the event horizon and is converted to the 
other wave packet of $k_{-s,\pm}$. We should note that the negative 
free fall frequency mode of $k_-$ comes to arise in the process of 
the mode conversion, while the other modes of $k_{-s,+}$ are 
positive ones.


The situation we consider in the figure \ref{fig-mcrad}, where no 
transmitted wave acrosses the event horizon, corresponds just to a 
boundary condition for the calculation of the original Hawking 
radiation. In the asymptotically flat region, $x \to \infty$, we can 
decompose the wave function as
  \sikib
    \Phi(x\! \to\! \infty) =
      \sum_{l=\pm s,\pm} c_l \exp(-i\om+k_l x) \, ,
    \label{eq-5}
  \sikie
where each coefficient, $c_{\pm s, \pm}$, can be uniquely determined 
by this boundary condition. Then we can obtain the form of the black 
hole radiation spectrum observed at the asymptotically flat region, 
$N(\om)$, by Bogoliubov transformation\cite{ref-CJ1}. It becomes
  \sikib
    N(\om) &=& \left< 0_{i} \right|
       a^{\dag}_{f} a_{f} \left| 0_{i} \right> \nonumber \\
     &=& \left| \frac{\omf(k_-) V_g(k_-) c_-^2(\om)}
                   {\omf(k_{+s}) V_g(k_{+s}) c_{+s}^2(\om)}
       \right| \, ,
    \label{eq-6}
  \sikie
where $\left| 0_{i} \right>$ is the vacuum state of the free fall 
observer at the initial time ($t \! \to \! -\infty $), 
$a^{\dag}_{f}$ and $a_{f}$ are the creation and annihilation 
operators of free fall observer at the final time 
($t \! \to \! \infty $). In deriving (\ref{eq-6}), we have assumed 
the Killing frequency spectrum of the wave packet, $\om(k)$, is 
sharply peaked around the four wave numbers $k_{\pm s,\pm}$.


\section{BH radiation specrtum in MC model}\label{sec-spectrum}

\subsection{Analytical calculation of the spectrum}

\subsubsection{The plan to calculate $N(\om)$}

We consider only the mode function corresponding to the wave 
number at the peak of $\om(k)$. We set its form as
  \sikib
    \Phi(t,x) = \exp(-i \om t) \, \px(x) \, ,
    \label{eq-7}
  \sikie
where $\om$ is the peak value of the Killing frequency spectrum 
of the wave packet. Then EOM (\ref{eq-3}) becomes
  \sikib
    \al \px^{(4)} &+& ( 1-v^2 ) \px'' + \nonumber \\
    && 2v(i\om - v') \px'
           + (\om^2 + i\om v')\px = 0 \, .
    \label{eq-8}
  \sikie
As mentioned at the end of previous section, we adopt the boundary 
condition of total reflection, which is expressed as
  \begin{description}
    \item[B.C.:] {\it The solution of} (\ref{eq-8})
                 {\it dumps with decreasing} $x$
                 {\it \\ inside the event horizon}.
  \end{description}
We take the origin of spatial coordinate $x$ at the location 
of the event horizon, that is, $v(x_{h}=0) = -1$. 

To calculate $N(\om)$, we follow the same plan as in \cite{ref-CJ2}. 
This plan consists of three steps: First step is to solve 
(\ref{eq-8}) around the event horizon. We use the Laplace 
transformation in this step, where the contour of the Laplace 
integral should be chosen as it satisfies the B.C. Secondly, we look 
for the solution of (\ref{eq-8}) far from the event horizon by means 
of the WKB approximation. We match the WKB solutions with those of 
the first step. The existence of the overlap region (matching zone) 
should be checked, where the solutions in first and second steps are 
valid. Third step is to calculate $N(\om)$ at the spatial infinity 
with using (\ref{eq-6}).

\subsubsection{Formulas for each step and small parameters}

We expand $v(x)$ around the event horizon as
  \sikib
    v(x) &\simeq& -1 + \ka x
    \, , \label{eq-9} \\
    1-v(x)^2 &\simeq& 2 \ka x \, , \nonumber
  \sikie
where we require $|\ka x| \ll 1$ and $\ka$ is the surface gravity 
which is assumed to be small itself. We use Laplace transformation 
of $\px(x)$,
  \sikib
    \px(x) = \int_C ds \, e^{sx} \ps(s) \, .
  \label{eq-10}
  \sikie
With approximating every coefficient in EOM (\ref{eq-8}) up to 
$O(\ka)$ , we obtain
  \sikib
    2 \ka f(s) \, \ps'(s) - B(s) \, \ps(s) = 0 \, , 
    \label{eq-11}
  \sikie
where $f(s) = s^2 + i \om s$, $B(s) = \al s^4 - \ka \be f'(s)$ 
and $\be = 1 + i \om / \ka$. This gives us
  \sikib
   &\px(x)& \nonumber \\
   &=& \int_{C} ds \,
           \left( s + i \om \right)^{i\al\om^3/2\ka}
           \left( s^2 + i \om s \right)^{-\be/2}
              \times \nonumber \\
   &&  \quad \exp \left[ sx + \frac{\al}{2 \ka} 
                          \left( \frac{1}{3} s^3
                               - i \frac{\om}{2} s^2
                               - \om^2 s - i \frac{11}{6}\om^3
                          \right)
                   \right] \nonumber \\
   &\equiv& \int_{C} ds \, \exp \left( h(s) \right)
   \label{eq-12} \, .
  \sikie
We will find later that the factor $(s+i\om)^{i\al\om^3/2\ka}$, 
which is ignored in \cite{ref-CJ2}, will make important contributions 
to the deviation of the MC spectrum from the thermal one. 

For WKB approximation, we set
  \sikib
    \px(x) = \exp \left[ i \int^x_{x_0} k(x') dx' \right] \, ,
  \label{eq-13}
  \sikie
where $x_0$ is a constant. We expand 
$k(x) = k_0(x) + (1/\la) k_1(x) + \cdots$ in introducing the scaling 
$x \to \la x$ ($\pdx \to (1/\la)\pdx$) \cite{ref-CJ2}, then we 
obtain from (\ref{eq-8}) that
  \sikib
    O(1) &:& \al k_0^4 - (1-v^2) k_0^2
                       - 2\om v k_0 + \om^2 = 0
      \label{eq-14} \\
    O(\la^{-1} ) &:& k_1 = \frac{i}{2}\frac{d}{dx} \ln
          \left[ \sqrt{\al} \left( \, 2\al k_0^3 \right. \right.
           \nonumber \\
    && \qquad\qquad\qquad  \left. \left.
        - (1-v^2) k_0 - \om v \, \right) \right]
      \label{eq-15}  \, ,
  \sikie
where four solutions of $k$ correspond to $k_{\pm,\pm s}$. 

To calculate $N(\om)$, we should evaluate $\omf V_g$. This is given 
by (\ref{eq-4}) to be
  \sikib
    \omf V_g = -2\al k_0^3 + (1-v_0^2)k_0 + \om v_0 \, .
  \label{eq-16}
  \sikie

As stated in section \ref{sec-intro}, $N(\om)$ deviates from the 
thermal spectrum in the high frequency region $\ka < \om$ 
\cite{ref-CJ1}. So we may not be able to derive the deviation with 
expanding $k_0$ around $\om = 0$ as done in \cite{ref-CJ2}. Thus we 
focus our attention on the region, $\ka < \om$, in order to derive 
the deviation. It is expected that the deviation becomes visible at 
the value of $\om$ which does not exceed $\ka$ so much. Thus, when 
we analyze how the MC spectrum deviates from the thermal one, the 
fourth term in the left-hand-side of (\ref{eq-14}) can be neglected 
in obtaining WKB solutions. This means to require 
  \sikib
    \al k_0^4 \gg \om^2 \, , \,
    (1-v^2)k_0^2 \gg \om^2 \, , \,
    \om v k_0 \gg \om^2 \, .
  \label{eq-17}
  \sikie
The other restrictions on the region of $\ka$ and $\om$ come from 
the validity of the solution of WKB and that of the Laplace 
transformation. These restrictions are expressed by $|k_1/k_0|\ll 1$ 
from WKB and by $\ka x \ll 1$ and $|s_0 x| \gg 1$ from Laplace 
transformation, where $s_0$ is a saddle point of integrand in 
(\ref{eq-12}) in using the asymptotic expansion of $x$ by the method 
of steepest descent contour (SDC). We can obtain the matching zone 
from these inequalities. It becomes 
  \sikib
    \left( \frac{\al}{\ka} \right)^{1/3} \ll x \ll \frac{1}{\ka}
    \, , \label{eq-18}
  \sikie
where we use the forms of $s_0$ and $k_{0,1}$ which are calculated 
later in appendix \ref{app-Laplace} and at (\ref{eq-23}). 
Furthermore, we require $\om x \ll 1$ for later convinience.

With introducing $\la$ and $\mu$ by $\ka x = (1/\ka \sqal)^{\la}$ and 
$\om = \ka (1/\ka \sqal)^{\mu}$ where $\ka \sqal \ll 1$, 
the parameter region we discussed above can be expressed as
  \sikib
    3\la>2\,(\mu-1) \, , \, -\frac{2}{3}<\la<0 \, , \,
    0<\mu \, , \, \la + \mu <0 \, , \label{eq-19}
  \sikie
where first and second inequalities obtained by (\ref{eq-17}) with 
(\ref{eq-9}) and (\ref{eq-18}) respectively.

\begin{figure}[t]
  \epsfysize=57mm
  \epsfbox{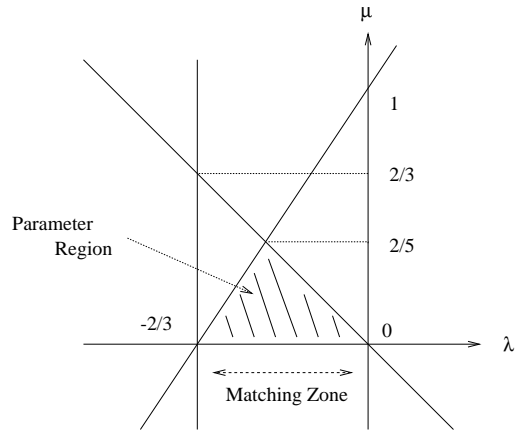}
  \caption{Parameter region we consider.}
  \label{fig-parameter}
\end{figure}

\subsubsection{Calculation of $N(\om)$}

We can proceed to calculate $N(\om)$ with above preparations. As we 
carry our the Laplace integral of (\ref{eq-12}) in the appendix 
\ref{app-Laplace}, we obtain the solution of (\ref{eq-8}) in the 
matching zone as
  \sikib
    \px(x) = \px_1(x) + \px_2(x) + \px_3(x) \, ,
    \label{eq-20}
  \sikie
where, as demonstrated below, $\px_{1,2,3}$ should be matched with 
WKB solutions, $\px_{\pm,+s}$, respectively. These are expressed as
  \sikib
   &\px_{1,2}(x)& \nonumber \\
   &=& \sqrt{\pi} \, 
        e^{i\pi(n+1/2-\ep3/4)-i11\al\om^3/12\ka} \times
        \nonumber \\
   &&  e^{\ep\pi\om/2\ka(1-\al\om^2/2)} \, 
        \left( \frac{2\ka}{\al}
        \right)^{-1/4-i\om/2\ka+i\al\om^3/4\ka} \times
        \nonumber \\
   &&  x^{-3/4-i\om/2\ka} \,
        \exp\left[ \ep i\frac{2}{3}
                        \sqrt{ \frac{2\ka}{\al} } x^{3/2}
            +i\frac{\om}{2} x \right]
        \, , \label{eq-21} \\
   &\px_3(x)& \nonumber \\
   &=& \, e^{-5\xi/3} \, 
        2 \, \sinh
          \left[
            \frac{\pi\om}{\ka}\left( 1-\frac{\al\om^2}{2} \right)
          \right] \times
        \nonumber \\
   &&  \Gamma \left[ -i \frac{\om}{\ka}
                 \left( 1-\frac{\al\om^2}{2} \right)
              \right] \,
        x^{i\om/\ka}
        \, , \label{eq-22}
  \sikie
where $(\ep,n) = (-1,0)$, $(1,1)$ for $\px_1$ and $\px_2$ 
respectively and $\xi = i\al\om^3/4\ka$. The factor $1-\al\om^2/2$ 
shows the difference form the analysis in \cite{ref-CJ2}, which arises  
from the factor $(s+i\om)^{i\al\om^3/2\ka}$ in the integrand of 
(\ref{eq-12}).

Next we shift our step to obtaining WKB solutions. We can obtain 
$k_0$ from (\ref{eq-14}) with the conditions of (\ref{eq-17}) 
only. Then $k_0$ becmoes
 \sikib
    k_0 \simeq
    \left\{
     \begin{array}{lcl}
      \om/(1+v)
         & \mbox{for} & k_{+s} \\
      \pm \sqrt{(1-v^2)/\al} + v\om/(1-v^2)
         & \mbox{for} & k_{\pm}
     \end{array}
  \right. \, ,
  \label{eq-23}
  \sikie
where these forms of $k_0$ are valid in the region $3\la >2(\mu -1)$ 
which is given by (\ref{eq-17}). Thus, for the purpose of matching 
with (\ref{eq-20}), we obtain WKB solutions in the parameter region 
of (\ref{eq-19}) as
  \sikib
    \px_{\pm}(x) &\simeq&
      A_{\pm} \, (\pm2\ka)^{-3/4} \, x^{-3/4 -i\om/2\ka} \times
      \nonumber \\
    && \quad
      \exp\left[ \pm\frac{2}{3}\sqrt{ \frac{2\ka}{\al} } x^{3/2}
                 + i\frac{\om}{2} x
          \right] \, ,
        \label{eq-24} \\
    \px_{+s}(x) &\simeq&
      A_{+s} x^{i\om/\ka}
      \label{eq-25} \, ,
  \sikie
where $A_{\pm,+s}$ are the coefficients obtained by $x_0$ in 
(\ref{eq-13}).

By matching WKB solutions, (\ref{eq-24}) and (\ref{eq-25}), with 
that of Laplace transformation,  (\ref{eq-20}), we obtain the wave 
function as
  \sikib
   &\px(x)& \nonumber \\
   &=& \, \sqrt{\pi} \, 
        e^{i9\pi/4-11\xi/3} \times
        \nonumber \\
   &&  \quad
       e^{-\pi\om/2\ka(1-\al\om^2/2)} \, 
        \left( \frac{2\ka}{\al}
        \right)^{-1/4-i\om/2\ka+\xi} \times
        \nonumber \\
   && \quad
      \left[ \, \frac{(-2\ka)^{3/4}}{A_-} \, \px_-(x) \right.
       \nonumber \\
   && \qquad\quad \left.
            + \, e^{-i5\pi/2 +\pi\om/\ka(1-\al\om^2/2)} \,
              \frac{(2\ka)^{3/4}}{A_+} \, \px_+(x) \,
      \right]
      \nonumber \\
   &+& \, \, e^{-5\xi/3} \, 
        2 \, \sinh
          \left[
            \frac{\pi\om}{\ka}\left( 1-\frac{\al\om^2}{2} \right)
          \right] \times
        \nonumber \\
   && \quad
      \Gamma \left[ -i \frac{\om}{\ka}
                 \left( 1-\frac{\al\om^2}{2} \right)
              \right] \,
        \frac{ 1 }{ A_{+s} } \, \px_{+s}(x)
   \, . \label{eq-26}
  \sikie
Note that we can use this result beyond the matching zone, i.e. 
$(\al/\ka)^{1/3}\ll x$ which is given by the validity of WKB 
approximaton, $|k_1/k_0| \ll 1$.

To obtain $N(\om)$, we need to evaluate $\omf V_g$ and $\px(x)$ 
at the asymptotically flat region. We obtain with taking the limit 
$v(x\to\infty)\,\to v_0$ that
  \sikib
    \px_{\pm}(x\to\infty)
    \simeq &&
      A_{\pm} \, (\pm1)^{-1/2} (1-v_0^2)^{-3/4} \times
        \nonumber \\
    &&  \quad \times
      \exp\left[ \pm i \sqrt{ \frac{1-v_0^2}{\al} } x
             + i \frac{v_0 \om}{1-v_0^2} x \right]
      \label{eq-27} \, , \\
    \px_{+s}(x\to\infty)
      \simeq &&
      A_{+s} \exp\left[ -i \frac{2v_0 \om}{1-v_0^2} x \right]
      \label{eq-28} \, ,
  \sikie
and
  \sikib
    \omf(k_{\pm}) V_g(k_{\pm}) &\simeq&
      \mp \frac{(1-v_0^2)^{3/2}}{\sqrt{\al}}
      \label{eq-29} \, , \\
    \omf(k_{+s}) V_g(k_{+s}) &\simeq& -\om
      \label{eq-30} \, .
  \sikie
Then we get $c_-$ and $c_{+s}$ by (\ref{eq-26}), (\ref{eq-27}) 
and (\ref{eq-28}), and finally obtain $N(\om)$ by (\ref{eq-6}) 
to be
  \sikib
    N(\om)
    = \frac{1-\al\om^2/2}
           {\exp[ 
            (2\pi\om/\ka) \,
            (1-\al\om^2/2)] - 1}  \, .
  \label{eq-31}
  \sikie
This result shows that MC spectrum is enhanced in comparison 
with Hawking spectrum. 

\subsection{Comparison with numerical results}

We compare the analytically derived form of the spectrum, 
(\ref{eq-31}), with numerical calculations. The numerical 
calculations are carried out with two types of $v(x)$, which are 
shown successively in the following two subsections.

We have used $Mathematica$ to solve (\ref{eq-8}) and fit its 
solution with the form, $\px = \sum_{l=\pm,\pm S} c_l \exp(ik_{l})$, 
in order to get $c_-$ and $c_{+s}$. 

Note that we have set $\ka/2\pi = 0.0008$ and $\al=1$ througout 
these numerical computations.

\subsubsection{First type of $v(x)$ -- CJ type}

First example of $v(x)$, which we call Corley-Jacobson (CJ) 
type\cite{ref-CJ1}, takes the form
  \sikib
    v(x) &=& {\rm sgn}(x) \, \frac{1}{2}
           \sqrt{ \tanh\left[ (2\ka x)^2 \right] } - 1
    \, . \label{eq-32}
  \sikie
To let $k_-$ mode be a negative free fall frequency one at the 
asymptotically flat region, the value of $\om$ is restricted by 
upper value $\om_{max} \sim 0.16$. Further, in our parameter region 
(\ref{eq-19}), the range of $\om$ is bounded as  
$0.005<\om<0.04 \, \, (0<\mu<2/5)$.

\begin{figure}[t]
  \epsfxsize=75mm
  \epsfbox{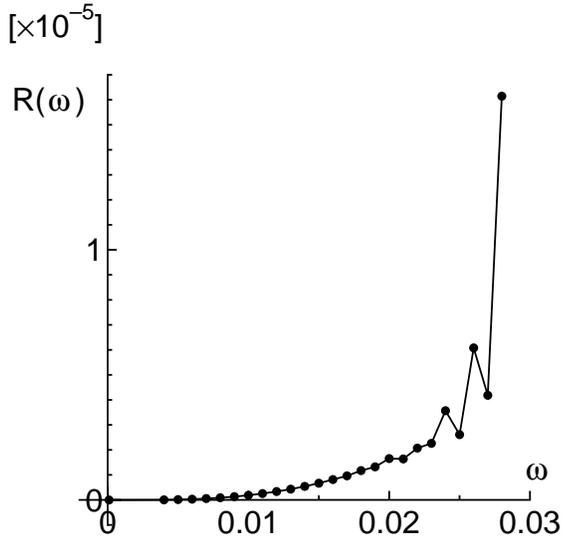}
  \caption{Graph of relative deviation, $R(\om)$, with CJ type of
           $v(x)$. $R(\om)$ grows qualitatively more and more with 
           increasing $\om$.}
  \label{fig-th2}
\end{figure}

We have calculated the relative deviation, 
  \sikib
    R(\om) = \frac{N(\om)}{N_{H}(\om)} -1 \, , \label{temp-4}
  \sikie
where $N_{H}$ is the Hawking spectrum. Figure \ref{fig-th2} is the 
plot of this relative deviation. $R(\om)$ grows qualitatively more 
and more with increasing $\om$.

This shows that the MC spectrum is enhanced in comparison with the 
Hawking spectrum, as is indicated by the analytical result 
(\ref{eq-31}). However figure \ref{fig-reldevlog}, which is the 
plots of the logarithm of relative deviations, exhibit that the 
order of $R(\om)$ in CJ type differs largely from our result 
(\ref{eq-31}). We expect that this difference can be reduced by 
changing the form of $v(x)$. The clue of this change is in the 
EOM (\ref{eq-8}). The most effective term for determining the 
analyticity of the solution of (\ref{eq-8}) is the second 
term, $(1-v^2) \px''$.
\footnote[5]{ This is also implied in section \ref{sec-SD}. } 
Thus, in order to compare (\ref{eq-31}) with a numerical 
calculation, the expansion $1-v^2 \simeq 2 \ka x$ near horizon 
should be better than the expansion $v \simeq -1 + \ka x$. 
By the way, $v(x)$ of CJ type can be expanded near horizon as:
  \sikib
    v(x) &\simeq& -1 + \ka x + O\left( (\ka x)^4 \right)
    \, , \label{temp-2} \\
    1-v(x)^2 &\simeq& 2 \ka x + O\left( (\ka x)^2 \right)
    \, . \nonumber
  \sikie
While CJ type is good form for the expansion $v \simeq -1 + \ka x$ 
near horizon, however it is not good one for the expansion 
$1-v^2 \simeq 2 \ka x$. We should modify $v(x)$ to let the 
expansion $1-v^2 \simeq 2 \ka x$ near horizon be better than CJ type.

\subsubsection{Second type of $v(x)$ -- DS type}

\begin{figure}[t]
  \epsfysize=65mm
  \epsfbox{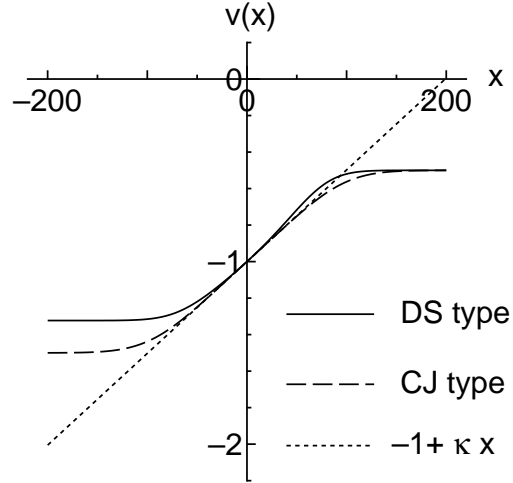}
  \caption{Shapes of $v(x)$ in the form of CJ and DS types. They 
           vary from $-1$ at $x=0$ to $x=-1/2$ at $x\to\infty$.}
  \label{fig-v}
\end{figure}

The modified type of $v(x)$, which is called as Double-Squareroot-Tanh 
(DS) type hereafter, is
  \sikib
    v(x) &=& -\sqrt{1 - {\rm sgn}(x) \frac{3}{4} \,
                  \sqrt{ \tanh\left[ \left(
                          \frac{8}{3}\ka x \right)^2 \right]
                  }}
    \, . \label{temp-1}
  \sikie
The range of $\om$ is $0.005<\om<0.04 \, \, (0<\mu<2/5)$, because of 
the same reason as that of CJ type. Figure \ref{fig-v} shows the 
shapes of $v(x)$ of CJ and DS types. They vary from $-1$ at $x=0$ to 
$x=-1/2$ at $x\to\infty$. 

The $v(x)$ of DS type takes the form near horizon:
  \sikib
    v(x) &\simeq& -1 + \ka x + O\left( (\ka x)^2 \right)
    \, , \label{temp-3} \\
    1-v(x)^2 &\simeq& 2 \ka x + O\left( (\ka x)^5 \right)
    \, . \nonumber
  \sikie
The expansion $v \simeq -1 + \ka x$ near horizon for CJ type is 
better than that for DS type, however the expansion 
$1-v^2 \simeq 2 \ka x$ for DS type is better than that for CJ type. 

\begin{figure}[t]
  \epsfxsize=75mm
  \epsfbox{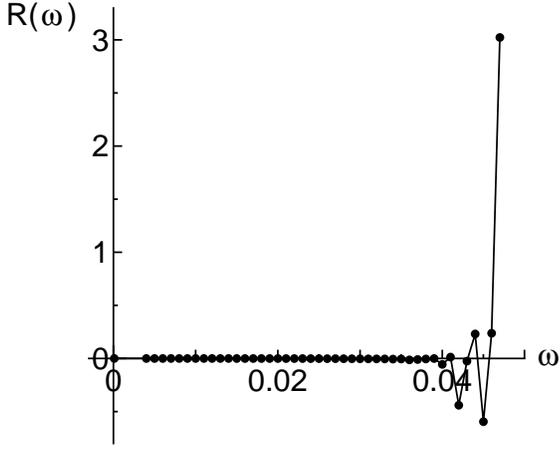}
  \caption{Graph of relative deviation, $R(\om)$, with DS type of 
           $v(x)$. $R(\om)$ grows qualitatively more and more with 
           increasing $\om$.}
  \label{fig-sqth}
\end{figure}

\begin{figure}[t]
  \epsfxsize=75mm
  \epsfbox{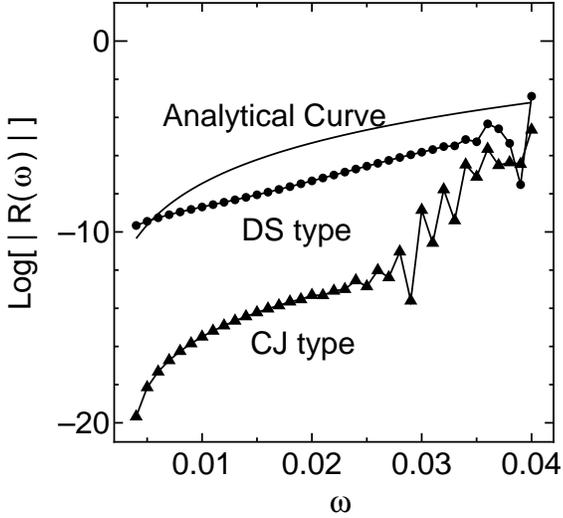}
  \caption{Graphs of logarithm of absolute value of the relative 
           deviations. The solid line is for our analytical result. 
           These show that our resultant spectrum explains the 
           qualitative nature of the MC spectrum.}
  \label{fig-reldevlog}
\end{figure}

Figure \ref{fig-sqth} is the plot of the relative deviation for DS 
type. $R(\om)$ grows qualitatively more and more with increasing 
$\om$. This figure and the figure \ref{fig-th2} implies that the 
details of the MC spectrum depend strongly on the form of $v(x)$. 
The total behavior of the figure \ref{fig-sqth}, however, shows 
that, althogh some data take negative value, the qualitative nature 
of the MC spectrum is enhancement in comparison with the Hawking 
spectrum, which is indecated by the analytical result (\ref{eq-31}). 
Then the figure \ref{fig-reldevlog} involves the plot of the 
absolute value of $R(\om)$ in DS type. This shows that the order of 
$R(\om)$ can be explained by our analytical result (\ref{eq-31}) 
fairly good, since the higher order effects by $1-v^2$ in DS type 
are weakened to agree with the assumption made in our analytical 
treatment of the MC spectrum. 

From above it is valid to mention that our resultant spectrum 
(\ref{eq-31}) can explain the qualitative nature of the MC spectrum 
in the range, $\ka < \om < \ka (\ka\sqal)^{-2/5}$.

\section{Summary and Discussion}\label{sec-SD}

We have derived the correction to the black hole radiation spectrum 
in the MC model in the frequecy range, $0<\mu < 2/5$, 
$\om = \ka (1/\ka\sqal)^{\mu}$. The MC model cut the high frequency 
mode off, so one may guess naively that the spectrum would be 
suppressed. Our result, however, shows that the MC spectrum is 
enhanced in the frequency range, $0<\mu < 2/5$. This is consistent 
with the numerical calculations as seen above.

The deviation factor $1-\al\om^2/2$ in (\ref{eq-31}) is understood 
as a correction of red shift which outgoing modes undergo during 
propagating from the event horizon to the asymptotically flat region. 
This is reflected on the form of $k_{+s}$ mode, 
$x^{i(\om/\ka)(1-\al\om^2/2)}$ in (\ref{app-6}). This form can be 
easily explained. Neglecting the fourth term in (\ref{eq-8}) which 
is of second order of small quantities $\ka$ and $\om$, we obtain
  \sikib
    \al\vpx''' + 2\ka x \vpx' - 2(1-\ka x)(i\om-\ka) \vpx = 0 \, ,
    \label{eq-33}
  \sikie
where we set $\px'(x)=\vpx(x)$ and $v(x)\simeq -1 +\ka x$. Under the 
assumption $\al < 1$, we expand 
$\vpx = \vpx_0 + \al \vpx_1 + \cdots$ which leads
  \sikib
    && 2\ka x \vpx_0' - 2(1-\ka x)(i\om-\ka) \vpx_0 = 0
       \label{eq-34} \\
    && 2\ka x \vpx_1' - 2(1-\ka x)(i\om-\ka) \vpx_1 = -\vpx_0'''
       \label{eq-35}
    \, .
  \sikie
The lowest order solution is given by
  \sikib
    \vpx_0(x) = x^{i\om/\ka -1} e^{(\ka-i\om)x} \, .
    \label{eq-36}
  \sikie
We are interested in the analyticity of the following form:
  \sikib
    \vpx_0 \sim x^{i\om/\ka -1}
      \quad \Rightarrow \quad
    \px_0 \sim x^{i\om/\ka} \, ,
    \label{eq-37}
  \sikie
since the temperature of the original Hawking radiation is determined 
by the form of an outgoing mode which takes the same form as above. 
Then we consider the correction to this analyticity. With setting 
$B=i\om/\ka -1$, we obtain $\vpx_1$ as 
  \sikib
    &\vpx_1(x)& \nonumber \\
    &=& -\frac{\al}{2}
         \left[ -\frac{B^3-3B^2+2B}{3\ka x^3}
                +\frac{3(B^3-B^2)}{2x^2}
         \right. \nonumber \\
    && \left. \qquad
                -\frac{3\ka B^3}{x} -\ka^2 B^3 \ln x
         \right] \, \vpx_0(x) \, .
    \label{eq-38}
  \sikie
We can recognize the fourth term in the square brackets as the 
secular one. Although this term is not dominant, it can change the 
analyticity which determines the temperature of the radiation. We 
use the renormalization group method in order to take all of the 
contributions of the secular term of $\vpx_1$ into account
\cite{ref-RG1}\cite{ref-RG2}. In the parameter region 
$\om/\ka \geq 1$, we obtain
  \sikib
    && \vpx \sim x^{\al(i\om-\ka)^3/2\ka +i\om/\ka -1}
            \sim x^{\al(i\om)^3/2\ka +i\om/\ka -1}
       \nonumber \\
    &\Rightarrow&
    \px \sim x^{i(\om/\ka)(1-\al\om^2/2)} \, ,
    \label{eq-39}
  \sikie
which shows the same deviation as in (\ref{eq-31}). 

Further, our result agrees with that of \cite{ref-HT}. In the paper 
\cite{ref-HT}, the order of the MC spectrum's deviation has been 
estimated, which is consistent with our spectrum, $N(\om)$.


\section*{Acknowledgements}

We would like to Thank J.Soda for his useful discussions. This work 
was supported in part by Monbusho Grant-in-Aid for Scientific 
Reserch No. 10640258. One of us (H.S.) also thanks S.Kawai for his 
helpful comments.


\appendix

\section{Calculation of Laplace integral}\label{app-Laplace}

We approximate (\ref{eq-12}) in this appendix by an asymptotic 
expansion of $x$ with the method of steepest descent contour (SDC). 
The saddle points, $s_0$, are the solutions of 
$dh/ds=0$ :
  \sikib
    \al s_0^4 + 2 \ka x s_0^2
    + 2 \left(i\om\ka x - \ka\be\right) s_0
    - i \om \ka \be = 0 \, . \label{app-1}
  \sikie
The $\px(x)$ approximated by SDC is expressed as 
  \sikib
    \px(x) &\simeq& \sqrt{2\pi} \,
      \left( |x|h''(s_0) \right)^{-1/2} \times \nonumber \\
    && \qquad
      \exp
      \left[ x h(s_0) +
        i \pi \left( \frac{1}{2}\gamma + n \right)
      \right] \, ,
  \label{app-2}
  \sikie
where $n$ is the integer determining the direction of the SDC and 
$\gamma = 0, \,1$ for $x<0$ and $x>0$ respectively. 

\begin{figure}[t]
  \epsfxsize=87mm
  \epsfbox{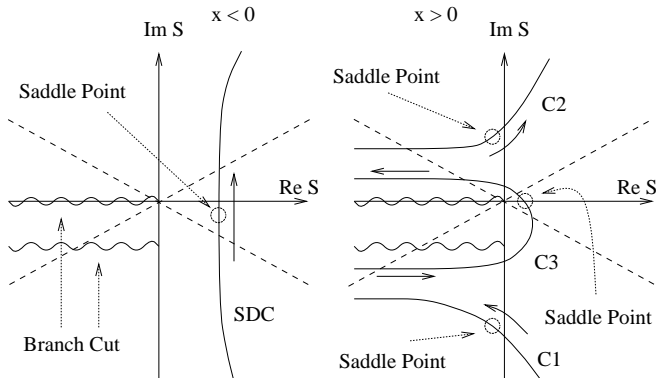}
  \caption{Contour of Laplace integral satisfying B.C.
           Left figure is for $x<0$. Right figure is for $x>0$.}
  \label{fig-SDC1}
\end{figure}

With approximating (\ref{app-1}) and (\ref{app-2}) in the paramater 
region (\ref{eq-19}), we obtain the saddle point satisfying B.C. in 
$x < 0$ as
  \sikib
    s_0 \simeq \sqrt{2\ka |x|/\al} - i\be/2|x| \, ,
  \label{app-3}
  \sikie
and for $x > 0$, the saddle points are
  \sikib
    s_0 \simeq
    \left\{
     \begin{array}{l}
      \be/x \\
      \pm i \sqrt{2 \ka x/\al} - \be/2x
     \end{array}
  \right. \, .
  \label{app-4}
  \sikie
The integral contour is determined as shown in the figure 
\ref{fig-SDC1}, where the contour should asymtpote to one of the 
three regions, $\pi/6 < \arg(s) < \pi/2$, 
$5\pi/6 < \arg(s) < 7\pi/6$ and $3\pi/2 < \arg(s) < 11\pi/6$ since, 
in the limit $|s|\to\infty$, the convergence of (\ref{eq-12}) 
requires $Re(s^3)<0$. The waving curves are the branch cuts of the 
integrand in (\ref{eq-12}).

The SDC of saddle point $\be/x$ approximates the contribution 
from the contour $C_3$. We can find, however, that this contribuiton 
is obtained by integral representations of some special functions 
more correctly than by the SDC approxiamtion. So we use SDC 
approximation only for the contributions from the contours $C_1$ 
and $C_2$. The dominant dependence on $x$ of these contributions are 
obtained by (\ref{app-2}) to give us (\ref{eq-21}). 

\begin{figure}[t]
  \epsfxsize=87mm
  \epsfbox{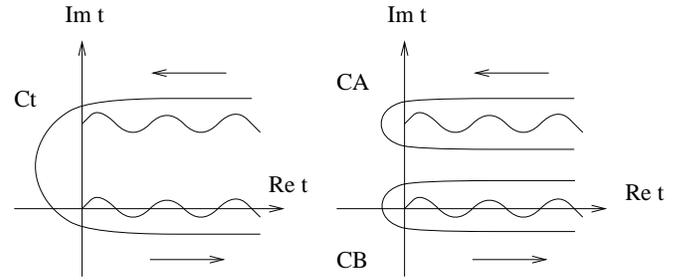}
  \caption{Contour of Laplace integral for $\px_3(x)$.}
  \label{fig-Ct}
\end{figure}

To carry out the contour integral of $C_3$, we transform $s$ to 
$t$ by $sx=e^{-i\pi}t$. While we choose the argument of $s$ as 
$-\pi<\arg(s)<\pi$, the argument of $t$ is to be 
$0<\arg(t)<2\pi$. Then (\ref{eq-12}) becomes
  \sikib
   &\px_3(x)& \nonumber \\
   &=& e^{-i\pi -i11\al\om^3/12\ka} \,
        (i\om)^{-\be/2 +i\al\om^3/2\ka} \, x^{\be/2 -1} \times
        \nonumber \\
   &&  \int_{C_t} dt \, (-t)^{-\be/2}
        \left( 1+ \frac{t}{-i\om x}
               \right)^{-\be/2 +i\al\om^3/2\ka} \times
        \nonumber \\
   &&  \exp
        \left[ -t + \frac{\al}{2\ka}
           \left( -\frac{t^3}{3x^3} -i\frac{\om t^2}{2x^2}
                  +\frac{\om^2 t}{x}
           \right)
        \right] \, ,
  \label{app-5}
  \sikie
where $C_t$ is the contour on $t$-plane corresponding to $C_3$, 
and $i\om = e^{i\pi/2} \om$ and $-i\om x = e^{i3\pi/2} \om x$. 
We can simplify the exponent of (\ref{app-5}) to be $\exp(-t)$ 
in our parameter region (\ref{eq-19}) by the same argument in 
\cite{ref-CJ2}: by $1/|\ka x^3| \ll 1$ and $\om x \ll 1$, the 
correction by 
$\exp[(\al/2\ka)( \cdots )] \simeq 1 + (\al/2\ka)( \cdots )$ is 
negligible. With dividing the contour $C_t$ into $C_A$ and $C_B$ 
as shown in the figure \ref{fig-Ct} and using the integral 
representation of Whittaker function, we can express (\ref{app-5}) 
as the sum of two Whittaker functions. 

Further we can simplify it by one confluent hypergeometric function 
to be
  \sikib
   &\px_3(x)& \nonumber \\
   &=& e^{-11\xi/3} \, 
        2 \, \sinh
          \left[
            \frac{\pi\om}{\ka}\left( 1-\frac{\al\om^2}{2} \right)
          \right] \times
        \nonumber \\
   && \quad
       \Gamma \left[ -i \frac{\om}{\ka}
                 \left( 1-\frac{\al\om^2}{2} \right)
              \right] \times
        \nonumber \\
   && \qquad
       e^{-i\om x} \, x^{i\om/\ka(1 -\al\om^2/2)} \times
        \nonumber \\
   && \qquad \quad
       F\left[ \frac{\be}{2}
             , 1 +i\frac{\om}{\ka}\left( 1-\frac{\al\om^2}{2} \right)
             ; i\om x
        \right] \, , \label{app-6}
  \sikie
where $\xi = i\al\om^3/4\ka$, $\be=1+i\om/\ka$, 
$\Gamma$ is the gamma function and $F$ is the confluent 
hypergeometric function. Then we obtain the dominant dependence on 
$x$ of $\px_3$, which bocomes (\ref{eq-22}).


\end{document}